\documentstyle[epsfig,psfig]{texas}

\begin{document}

\def\apj{ApJ}
\def\keV{\hbox{keV}}
\def\MeV{\hbox{MeV}}
\def\s{\hbox{s}}

\title{The Role of the BATSE Instrument Response in
Creating the GRB E-Peak Distribution}

\author{J. J. Brainerd,$^{1}$
G. Pendleton,
R. Mallozzi,
M. S. Briggs,
R. D. Preece}
\address{University of Alabama in Huntsville\\
(1) E-mail: Jim.Brainerd@msfc.nasa.gov
}

%
%

\begin{abstract}
All gamma-ray bursts are observed to have approximately the same
characteristic gamma-ray energy.  We show in this article that
for bursts in the BATSE data set, this property as measured by the E-peak
value is not an instrumental effect, but a physical
property of gamma-ray bursts.
\end{abstract}

%
%

\section{The Characteristic Photon Energy of Bursts}

The prompt emission of gamma-ray bursts is observed
predominately at several hundred keV.  This is one
of the most interesting features of gamma-ray bursts,
since it is not easily explained by most theories of prompt
gamma-ray emission.  In particular, the internal and
external shock theories predict a wide variation in
the characteristic gamma-ray energy both
during a burst and between bursts.
It is therefore imperative to know whether this is a
physical characteristic of bursts, or a consequence of
an instrumental effect.

In this article, we discuss the instrumental effects
that arise when observing gamma-ray bursts with
the Burst and Transient Source Experiment (BATSE) on the Compton
Gamma-Ray Observatory (CGRO).\cite{brainerd}
The primary points that
affect the observed
distribution function are the ability to correctly
define the characteristic energy of the gamma-ray burst and
the ability to trigger on a gamma-ray burst with
a characteristic energy outside of the trigger energy range.
These effects are discussed in \S 2 and \S 3 below.  Models
of simple distributions for the characteristic energy are
given in \S 4, and their consistency with the observations
is discussed in \S 5.  We find that the observed distribution
of characteristic energies cannot be explained through
instrumental effects alone.   A model distribution with a
characteristic energy is fit to the observations to quantify
the type of physical theory required by the observations.
Our conclusion is that the existence
of approximately the same characteristic energy for all gamma-ray
burst spectra is a physical property of gamma-ray bursts that any
viable theory must explain.

\section{Defining E-Peak}

One can characterize the gamma-ray burst spectrum through the
E-peak value ($E_p$), the photon energy at which the $\nu F_{\nu}$ curve
has a maximum.  The distribution of such values as measured by the
BATSE gamma-ray burst instrument is narrowly distributed.\cite{mallozzi}
The standard
method of modeling a gamma-ray spectrum is forward fitting.  In this
method, a model photon spectrum is folded through a model of the
detector response matrix (DRM) to produce a count spectrum.  The count
spectrum is then compared to the observed spectrum, with the best
fit found through $\chi^2$ minimization.  The photon model used in
the analysis of the BATSE gamma-ray burst spectra is the gamma-ray burst
spectral form,\cite{band} which is a 4 parameter model that produces
good fits  to the data.  The data type used in deriving $E_p$ for
the BATSE data set is the MER data type, which covers the
spectrum with 16 channels.

To test the ability of BATSE to correctly determine the value of
$E_p$, we generated test burst count spectra with background counts added
for model spectra of a given $E_p$, and
then went through the procedure of subtracting background and deriving
the values of $E_p$ that provide the best fit to the test spectra.
We find that the forward fitting
method correctly defines the value of $E_p$ for
$20 \, \keV < E_p < 2 \, \MeV$.  Once one is outside this range,
either no value of $E_p$ is found, or the value that is found is
at one of these two limits on $E_p$.  Because few BATSE bursts have
$E_p$ at $\approx 2 \, \MeV$, and no BATSE burst is at $\approx 20 \, \keV$,
the misidentification of the value of $E_p$ for $E_p > 2 \, \MeV$ and
$E_p < 20 \, \keV$ is inconsequential.

\begin{figure}
\centering
\epsfig{figure=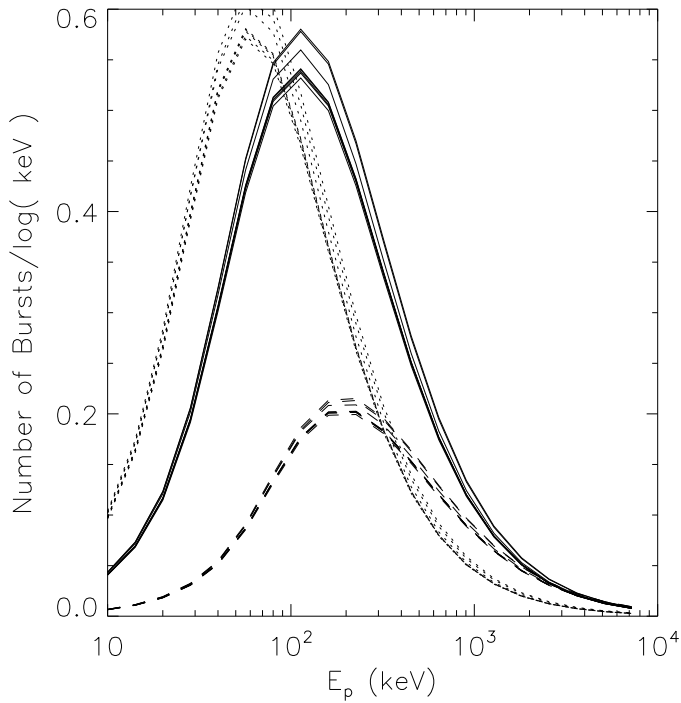, width=4.0in}
\caption{Expected E-Peak distribution for bursts
with $C > C_{min}$ such that most bursts lie on the $-3/2$ portion
of the cumulative peak-flux distribution.
Here, $C$ is the burst
count rate in the energy range of the trigger and $C_{min}$ is
the limiting count rate for selecting bursts.  The solid
curves are for triggering on an energy range of $50$--$300 \, \keV$,
the dotted lines is for triggering on a range of $20$--$100 \, \keV$,
and the dashed lines are for triggering on the range of $> 100 \, \keV$.
The different curves are for different burst directions relative to Earth.
The spectral model used in these curves is a grb spectral form with
$\alpha = -1$ and $\beta = -3$.
}
\end{figure}

\begin{figure}
\centering
\epsfig{figure=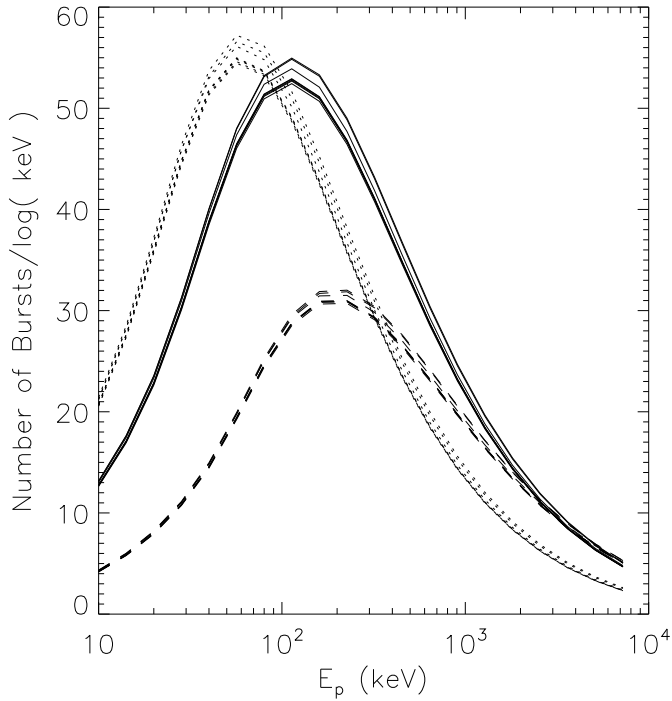, width=4.0in}
\caption{Expected E-Peak distribution for bursts with
$C > C_{min}$, with $C_{min}$ selected to place most bursts
on the $-0.8$ portion of the cumulative peak-flux
distribution.  The remainder of the free parameters in the models is
as in Fig.~1.
}
\end{figure}

\section{BATSE Triggering}

The BATSE instrument triggers on a gamma-ray burst when the count rate
in the trigger channels exceed an average background count rate by
a preset value.  Of the eight modules that comprise BATSE, at least two
must be above the background rate for the burst to trigger.
The triggering is done using four discriminator channels on
the timescales of $1.024 \, \s$, $0.256 \, \s$, and $0.064 \, \s$.
For most bursts, the triggering is on
channels $2 + 3$, which contain counts in the $50 \, \keV$ to
$300 \, \keV$ energy range.

To model the trigger of a gamma-ray burst, we ran Monte Carlo simulations
of the count rates generated by a gamma-ray burst of a given normalization
and value of $E_p$.  The randomness in this simulation was in the
orientation of the spacecraft relative to the gamma-ray burst and to
the center of the earth.  Two aspects that affect the count rate
found for a burst are the orientation of all detectors to the burst, and
the component of the burst scattered into the detectors from Earth's
atmosphere.  One aspect this simulation did not address was the
angular dependence of the background count rate.  The purpose of the
analysis was to determine the average count rate in the second brightest
detector as a function of $E_p$.  The result is used to
relate the photon rate of a burst to the count rate in the trigger
channels.

\begin{figure}
\centering
\epsfig{figure=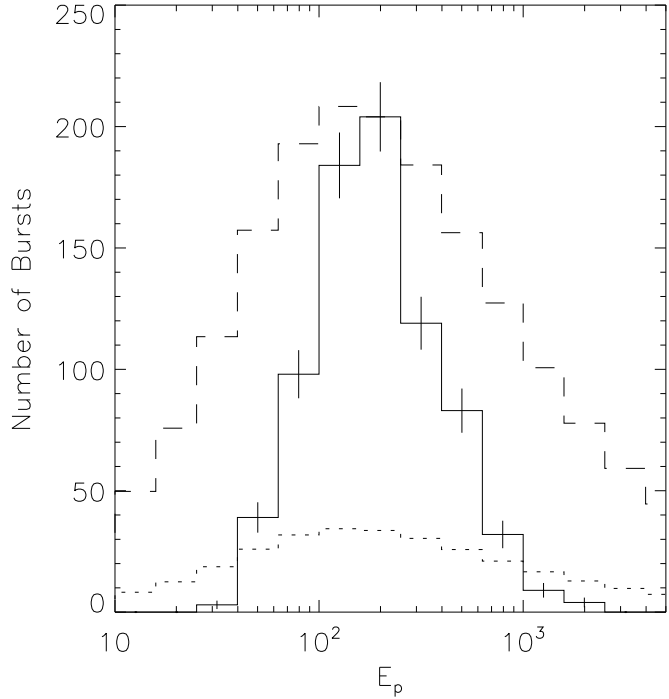, width=4.0in}
\caption{Best fit of power law E-peak distribution to
the BATSE E-peak distribution.  The solid curve gives the
observed distribution of $E_p$.  The dotted curve is the best
fit model, while the dashed curve is the best fit model renormalized
to match the maximum of the observed distribution.
}
\end{figure}

\begin{figure}
\centering
\epsfig{figure=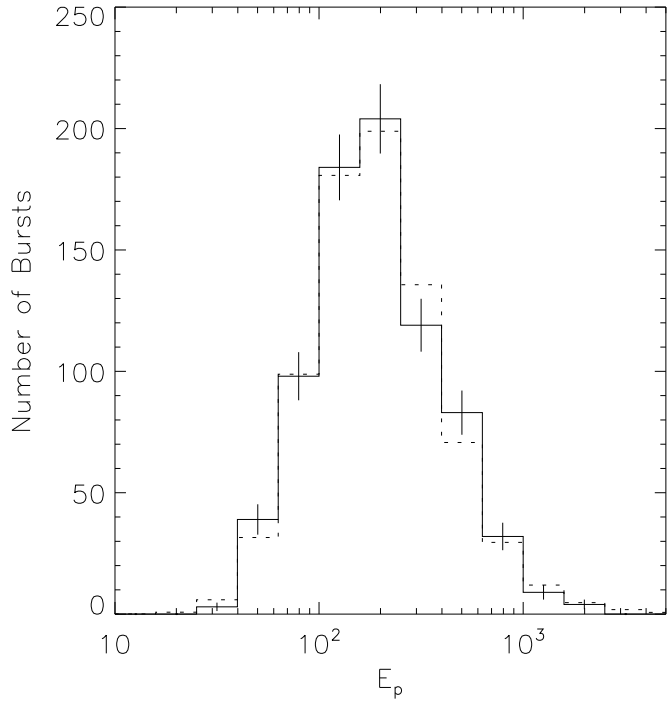, width=4.0in}
\caption{Best fit of log normal E-peak distribution to
the BATSE E-peak distribution.  The solid curve is the observed
distribution, while the dotted curve is the best fit model.
}
\end{figure}

\section{Model E-Peak Distributions}

Based on the conversion of photon spectrum to count rate, one can
convert model distributions  in $E_p$ and $N_p$, the photon rate
at $E_p$, into distributions in $E_p$ as functions of $C_{min}$,
the minimum count rate.  In doing this, we assume that
the minimum count rate is high enough above the background that
threshold effects are unimportant.  We are also ignoring the fact
that different detectors will have different background count rates.
We assume that the distribution of $N_p$ is a broken power law similar
to the peak-flux broken power law, with the power law having an index
of $-1.8$ below the break, and an index of $-5/2$ above the break.
The value at which power law index breaks in the $N_p$ distribution
is assumed to depend on $E_p$ as $E_p^{-2}$, which
is equivalent to the statement that the total energy per unit time emitted
as gamma-rays is independent of $E_p$ for bursts that fall on the break
of the peak-flux curve.  Two models for the $E_p$ distribution are
used.  The first is a power law, so that there is no characteristic
value of $E_p$ defined by the physics.  The second is a log normal
curve with two power laws joined on to the wings of the distribution.
This second is used to fit the observed $E_p$ distribution to demonstrate the
extent to which the physical $E_p$ distribution must break to reproduce
the observations.

The model distributions for a power law $E_p$ distribution are given
in Figures 1 and 2.  These curves are calculated for three different
trigger energy ranges: channels $1 + 2$ ($20 \, \keV$--$100 \, \keV$),
channels $2 + 3$ ($50 \, \keV$--$300 \, \keV$), and
channels $3 + 4$ ($> 100 \, \keV$).  The power law index for the $E_p$
distribution is set to $-1$.  The difference between Figures 1 and 2 are
in the minimum count rate.  The first has a minimum count rate such
that most bursts
are in the $-3/2$ portion of the cumulative peak-flux distribution,
while the second has a count rate such that most bursts are on
the $-0.8$ portion of the cumulative peak-flux distribution.
The effect of lowering the count rate limit is to broaden the
$E_p$ distribution.

\section{Fits of E-Peak Distributions}

The observed $E_p$ distribution was fit to the power law E-peak
distribution.  Most bursts have low count rates, placing them
below the break in the peak-flux distribution, so the low count rate
model for the $E_p$ distribution is used.
We find that the best power law index for the $E_p$
distribution is $-0.94 \pm 0.11$, with $\chi^2 = 444.4$ for 8 degrees
of freedom.  This is clearly a poor fit to the data.  This model and the
observed $E_p$ distribution are given in Figure 3.  The model distribution
is much broader than the observed distribution.  This demonstrates that the
physical distribution of $E_p$ must have a characteristic value near
$200 \, \keV$ to produce the observed distribution.

Just how narrow the physical distribution must be is demonstrated by the
fit of a log normal distribution to the observations.  In this distribution,
the half maxima are a factor of 4 apart, and the power law tails have
indexes of $3.35$ and $-2.58$, so that the change in index exceeds 5.
The model fit has $\chi^2 = 7.93$ for 5 degrees of freedom, which is
an excellent fit to the data.  The physical distribution therefore
must be very narrow to reproduce the observations.
This fit also suggests that above $1 \, \MeV$, each decade in $E_p$
contains $< 10 \%$ of the bursts in the lower decade, and that
below $100\, \keV$, each decade contains $< 1 \%$ of the bursts in the
higher decade.

\section{Discussion}

We have presented the results of our study of the BATSE instrumental
effects that affect the determination of the $E_p$ distribution.
Our results are as follows:

\begin{itemize}

\item Model fitting of spectra to gamma-ray burst spectra correctly
gives the value of $E_p$ for $20 \, \keV < E_p < 2 \, \MeV$.

\item The power law model curves for the $E_p$ distribution
for bursts that trigger on the $50$--$300 \, \keV$ count energy range
have maxima at $\approx 100 \, \keV$, and
half-maxima at $\approx 20 \, \keV$ and $\approx 800 \, \MeV$.

\item The power law model curves provide poor fits to the observations,
because the observations are at $\approx 10 \%$ of the peak values
at the energies where the model curves have their half-maxima. 

\item Fitting a log normal $E_p$ distribution to the observations finds
that the best fit is very narrow, with power law wings that differ
by 5 in index.

\end{itemize}

These results demonstrate that the narrow $E_p$ distribution must have
a physical origin.  Theories of the prompt gamma-ray emission must provide
an explanation for this property before they can be regarded as viable.

\section*{Acknowledgments}

This research was supported under the NASA grant NAG5-6746.

\end{document}